%% file: atos.tex
\documentclass[sigplan,10pt,nonacm,anonymous=false]{acmart}
\newcommand{\final}{1}
\usepackage{listings}
\usepackage[newfloat,frozencache=true,cachedir=./minted-atos]{minted}
\usepackage{algorithm}
\usepackage{algorithmicx}
\usepackage[noend]{algpseudocode}
\setminted{fontsize=\scriptsize}
\input{tex/macros}
\usepackage{array}
\newcolumntype{C}[1]{>{\centering\arraybackslash}p{#1}}
\usepackage{graphicx}
\usepackage{multirow}
\usepackage{siunitx}
\usepackage{booktabs}

\AtBeginDocument{%
  \providecommand\BibTeX{{%
    \normalfont B\kern-0.5em{\scshape i\kern-0.25em b}\kern-0.8em\TeX}}}

\newenvironment{code}{\captionsetup{type=listing}}{}
\SetupFloatingEnvironment{listing}{name=Listing}

\newfloat{lstfloat}{htbp}{lop}
\floatname{lstfloat}{Listing}

\acmConference[PPoPP 2022]{the 27th Symposium on Principles and Practice of Parallel Programming}{February 12--16, 2022}{Seoul, South Korea}
\acmYear{2022}
\acmISBN{} 
\acmDOI{} 
\startPage{1}

\setcopyright{acmcopyright}

\usepackage{booktabs}   
\usepackage{subcaption} 
\hypersetup{draft}

\begin{document}

\title{Atos: A Task-Parallel GPU Dynamic Scheduling Framework for Dynamic Irregular Computations}         


\author{Yuxin Chen}
\authornotemark[1]
\affiliation{%
  \institution{University of California, Davis}
}
\email{yxxchen@ucdavis.edu}

\author{Benjamin Brock}
\affiliation{%
  \institution{University of California, Berkeley}
}
\email{brock@cs.berkeley.edu}

\author{Serban Porumbescu}
\affiliation{%
  \institution{University of California, Davis}
}
\email{sdporumbescu@ucdavis.edu }

\author{Ayd\i n Bulu\c{c}}
\affiliation{%
  \institution{Lawrence Berkeley National Laboratory}
  }
\email{abuluc@lbl.gov}

\author{Katherine Yelick}
\affiliation{%
  \institution{University of California, Berkeley}
  }
\email{yelick@berkeley.edu}

\author{John D. Owens}
\affiliation{
\institution{University of California, Davis}
}
\email{jowens@ece.ucdavis.edu}

\renewcommand{\shortauthors}{Yuxin Chen, et al.}
\begin{abstract}
We present Atos, a task-parallel GPU dynamic scheduling framework that is especially suited to dynamic irregular applications.
Compared to the dominant Bulk Synchronous Parallel (BSP) frameworks, Atos exposes additional concurrency by supporting task-parallel formulations of applications with relaxed dependencies, achieving higher GPU utilization, which is particularly significant for problems with concurrency bottlenecks. Atos also offers implicit task-parallel load balancing in addition to data-parallel load balancing, providing users the flexibility to balance between them to achieve optimal performance.
Finally, Atos allows users to adapt to different use cases by controlling the kernel strategy and task-parallel granularity.
We demonstrate that each of these controls is important in practice.

We evaluate and analyze the performance of Atos vs.\ BSP on three applications: breadth-first search, PageRank, and graph coloring. Atos implementations achieve geomean speedups of 3.44x, 2.1x, and 2.77x and peak speedups of 12.8x, 3.2x, and 9.08x across three case studies, compared to a state-of-the-art BSP GPU implementation. Beyond simply quantifying the speedup, we extensively analyze the reasons behind each speedup.  This deeper understanding allows us to derive general guidelines for how to select the optimal Atos configuration for different applications.  Finally, our analysis provides insights for future dynamic scheduling framework designs.
\end{abstract}



\maketitle

\input{tex/intro}

\input{tex/bsp-ndp}
\input{tex/design_decisions}

\input{tex/framework_api}
\input{tex/graph-problems}
\input{tex/analysis}
\input{tex/conclusion}


\bibliographystyle{ACM-Reference-Format}
\bibliography{ics-async}



\end{document}

%% file: tex/macros.tex
\usepackage{xcolor}
\usepackage[prependcaption]{todonotes}
\presetkeys{todonotes}{author=TODO, inline, color=red!60}{}

\newcommand{\yuxin}     [1]{\todo[inline,author=Yuxin,color=green!20]{#1}}
\newcommand{\john}      [1]{\todo[inline,author=John,color=blue!20]{#1}}
\newcommand{\ben}      [1]{\todo[inline,author=Ben,color=red!20]{#1}}
\newcommand{\kathy}      [1]{\todo[inline,author=Kathy,color=violet!20]{#1}}

\newcommand{\intodo}    [1]{~\textbf{\color{red}TODO\@: #1}}
\renewcommand{\final}{1}
\ifthenelse{\equal{\final}{1}}
{
  \renewcommand{\todo}[1]{}
  \renewcommand{\yuxin}[1]{}
  \renewcommand{\john}[1]{}
  \renewcommand{\ben}[1]{}
  \renewcommand{\kathy}[1]{}

  \renewcommand{\intodo}[1]{}
}{}

%% file: tex/intro.tex
\section{Introduction}\label{sec:intro}
Bulk-synchronous parallel (BSP) programming~\cite{Valiant:1990:ABM} is the traditional model for GPU applications. It is a natural fit for statically schedulable, regular problems, such as many dense matrix, image analysis, and structured grid computations.  Programming environments like NVIDIA's CUDA and Khronos's SYCL support this relatively simple model, which maps efficiently to the massive fine-grained parallelism on GPUs and can deliver near-peak performance.

However, some important problems are instead irregular, with frequent control flow branches, non-unit stride memory accesses, variable amounts of work across loop iterations, and dynamically varying degrees of parallelism.  Algorithms that operate on graphs or trees or those with recursive formulations often exhibit such irregularity.  These more naturally use a task-based programming model.  Programming systems like Legion~\cite{bauer2012legion}, PTask~\cite{rossbach2011ptask} and StarPU~\cite{augonnet2010data} use tasking on the CPU to feed GPUs with kernels to mask the latency of communication and keep the GPU busy.
In contrast, we consider the problem of very fine-grained tasking where a set of similar application level tasks are aggregated to form a data-parallel GPU task. This idea is used in state-of-the-art GPU graph libraries like Gunrock~\cite{gunrock2017}, where the application level tasks are vertices or edges. Each frontier in a graph sweep is launched as a separate GPU kernel in the BSP model; this raises issues of insufficient parallelism, uneven finish times, and high kernel launch overhead for small frontiers.

To address the above issues, we present the design of Atos, a task-scheduling framework for GPUs that is adaptable to different usage scenarios:
\begin{itemize}
\item It supports both expensive and inexpensive frontiers by providing persistent and non-persistent task schedulers.  The persistent scheduler is a GPU kernel that runs continuously to minimize launch overhead.

\item It allows the user to trade off task and data parallelism by selecting the worker size, which is the number of GPU threads within each worker, and the number of items in each task.

\item It uses a single shared task queue, which balances load more quickly than a distributed queue, yet is fast enough to keep GPU workers occupied.

\item It supports asynchronous execution across frontiers to maximize available parallelism, while mostly preserving cross-frontier ordering and thus minimizing overwork.
\end{itemize}

We study three graph algorithms on a variety of graph datasets that stress the importance of each of the above controls.  These problems have nested parallelism with outer loop dependencies, and the ability to relax those dependencies comes at the cost of possible overwork. Thus a second major theme in this paper is this tradeoff between increased parallelism and overwork.  Each of our three algorithms explores this tradeoff in a somewhat different manner and detailed discussion and analysis is presented in Section~\ref{sec:graph-problems} and~\ref{sec:analysis}.

Our contributions include:
\begin{itemize}
  \item Developing a generalized GPU task-parallel framework that explores a broad design space of possible task-parallel implementations;
  \item A demonstration of the benefits of mixed task and data parallelism for fine-grained parallel applications;
  \item Identifying relaxed-synchronization applications as a strong candidate for acceleration with a GPU task-parallel framework; and
  \item A detailed analysis of application performance that highlights the impact of design decisions both within the task scheduler and at the application level.
\end{itemize}

%% file: tex/bsp-ndp.tex
\section{A Dynamic, Irregular Application Pattern}
\label{sec:problem-class}
Atos handles a broad set of applications with fine-grained task and data parallelism, but we choose to focus here on a particularly challenging class of irregular nested loops with the following form:
\begin{lstfloat}
  \begin{code}
    \begin{minted}
      [
      % frame=lines,
      framesep=2mm,
      baselinestretch=1.2,
      fontsize=\footnotesize,
       baselinestretch=1,
      % fontsize=8,
      xleftmargin=1em, linenos
      ]
      {cuda}
in_frontier = initialize()
while (stop condition not met):  // outer loop
  ---------------CUDA-KERNEL------------------
  for (i in in_frontier):  // inner loops
    for (j = 0 to workload(i).size()):
      out_frontier.append(f(in_frontier[i], j))
  ---------------------------------------------
  cudaDeviceSynchronize()
  in_frontier = out_frontier
    \end{minted}
    \captionof{listing}{A program with nested loops, expressed with a frontier abstraction.}
    \label{lst:nested_loop}
  \end{code}
\end{lstfloat}


The inner loops produce data parallelism that may be flattened as in NESL~\cite{Blelloch:1990:CCL} or Gunrock~\cite{gunrock2017} to maximize parallelism and to implement a data-parallel load-balancing technique.  For example, if the inner loops are iterating over graph vertices and outgoing edges, some pre-analysis may be used to evenly distribute the edges rather than the vertices.  Our work also relaxes the outer loop iterations, which will expand opportunities to find parallelism.

Our applications exhibit one or more of the following forms of dynamic, irregular parallelism:

\begin{itemize}
  \item \emph{The number of tasks varies across outer loop frontiers:} Work is generated dynamically and the number of output tasks produced from each input task is not fixed.

  \item \emph{The cost of each inner loop task (lines 4--5 in Listing~\ref{lst:nested_loop}) may vary:} The loop bound (\verb|workload(i).size()|) in line 5 is not fixed.

  \item \emph{Total work may vary:}  The outer loop has a loop-carried dependence, so while dynamic generation of tasks will enforce some dependencies, parallel execution of tasks across frontiers can change program behavior, including the total number of tasks processed.

\end{itemize}

\subsection{Performance Challenges of Dynamic, Irregular Problems}
\label{sec:bsp-performance-challenges}

Traditional BSP implementations of such applications usually launch a series of kernels, with each kernel corresponding to a bulk-synchronous step (one iteration of the outer loop). Between each step is a global synchronization barrier. The kernels themselves parallelize over the inner loops, with the list of work passed between iterations forming the ``frontier''.
While optimized BSP implementations of many dynamic irregular applications achieve impressive performance, we identify three performance challenges with this approach.

\textbf{Small frontier problem} The runtime of a kernel is the sum of fixed costs (the cost of the global synchronization barrier plus the kernel launch cost) and the amount of time to process the input frontier. The size of input frontiers across iterations may vary significantly over the lifetime of the program. A small frontier has two performance consequences: (1)~The fixed costs dominate the overall processing cost; the GPU is spending a significant amount of time setting up or waiting for computation rather than performing it. (2)~A small frontier may not be large enough to fill the GPU with work, leaving processing units idle~\cite{Gunrockfrontier}.

\textbf{Load imbalance} Because of the irregular nature of work in the inner loop, efficiently assigning work to GPU threads is critical to achieve the highest performance. Statically computing work assignment is infeasible because work is produced dynamically, so lines 4--6 in Listing~\ref{lst:nested_loop} are usually implemented with data-parallel load balancing techniques~\cite{merrillbfs,davidsonsssp, gunrock2017} that compute work assignment at runtime. However, the optimal load-balancing technique is prob\-lem-de\-pen\-dent and even for a fixed problem may be input-dependent. And the runtime cost of that technique may be high.


\textbf{Loss of concurrency opportunities} The BSP model enforces an ordering between every operation in iteration $i$ with every operation in iteration $i+1$, even if some of those pairs may be independent.  We relax this cross-frontier ordering in two ways. One is to consider problems whose typical formulation is as nested loops but whose computation is \emph{amenable to reordering} across BSP iterations. Two is to \emph{speculate} that work in later iterations is not dependent on work in earlier iterations and to repair the misspeculation or retry the computation if it is incorrect.

\kathy{Sorry, I don't what the last two sentences are trying to say.  The use of algorithm here is not defined -- as opposed to the application?  the code? Also, as a more general point if we want to talk about speculative execution it should come up earlier, but normally that refers to work that will be rolled back, which we are not doing.}

\john{Replaced ``algorithm'' with the vaguer but accurate ``computation''. Added ``retry the computation'' for the second one. I don't think ``repair'' implies ``rollback''. For BFS, we are ``repair''ing an incorrectly set depth by simply overwriting the incorrect depth with the correct one (or at least a more correct one).}

%% file: tex/design_decisions.tex
\vspace{-0.5\baselineskip}
\section{Our Task-Parallel Programming Model and Design Space}\label{sec:design_decisions}

Our programming model allows implementations of task-parallel formulations of the workloads we discussed in Section~\ref{sec:problem-class}, with a focus on  providing solutions to BSP's three challenges: small-frontier, load-imbalance, and loss of concurrency opportunities.
We use the following terminology:
\begin{description}
  \item[worker:]  one or a group of GPU threads.
  \item[task:] one or more pieces of work that are scheduled as a single unit in our system. 
  \item[application function $f()$:] the code that processes each task. Each application function declares the worker size it requires to run.
\end{description}
\begin{lstfloat}
  \begin{code}
    \begin{minted}
      [
      % frame=lines,
      framesep=2mm,
      baselinestretch=1.2,
      fontsize=\footnotesize,
      xleftmargin=1em,linenos,
      baselinestretch=1,
      % fontsize=8,
      % linenos
      ]
      {octave}
--------CUDA_KERNEL----------
for each worker:
  while not queue.empty():
    task = queue.concurrent_pop(task.size())
    new_tasks = f(task)
    queue.concurrent_push(new_tasks)
-----------------------------
    \end{minted}
    \captionof{listing}{SPMD code of each thread worker in Atos.}
    \label{lst:tp}
  \end{code}
\end{lstfloat}

At a high level, our model maintains a queue of tasks. Workers fetch a task from the queue, process the task, and add newly generated tasks (if any) to the queue. The program runs until either a stop condition is met or the queue is empty (Listing~\ref{lst:tp}). This programming model addresses the performance challenges from Section \ref{sec:bsp-performance-challenges}:
\begin{itemize}
  \item We can implement Listing~\ref{lst:tp} with a single kernel invocation, avoiding multiple launches of small kernels.
  \item Task parallelism is implicitly load-balanced because workers can run independently and stay busy even if tasks require different amounts of work.
  \item Listing~\ref{lst:tp} has no global synchronization; instead the programmer controls the scheduling of work, allowing more flexible dependencies and thus more opportunities for concurrency.
\end{itemize}

The primary focus of this paper is how to best implement Listing~\ref{lst:tp} on the GPU\@. We identify below the four key design decisions in such an implementation. 

\begin{description}
  \item[Relaxing barriers] A BSP implementation separates work in each iteration with a global barrier. Can we benefit from relaxing this global barrier constraint?
  \item[Worker size] We can choose what GPU resources we assign to each worker. What is the worker size that yields the best performance?
  \item[Data vs.\ Task Parallelism] We expect to leverage the parallelism between tasks. We can also choose the size of our tasks, and within each task, potentially leverage data parallelism. What is the right balance between task and data parallelism?
  \item[Kernel strategy] Listing~\ref{lst:tp} is written in a ``persistent'' style, implementable with a single kernel call. We could alternatively interchange its outer and inner loops, making one ``discrete'' kernel call per iteration. Which is best for optimal performance?
\end{description}

\subsection{Relaxing Barriers}

As discussed in Section~\ref{sec:problem-class}, many applications have the nested loop structure from Listing~\ref{lst:nested_loop} and their BSP implementations may lose concurrency opportunities because of the global barrier between the outer loops.
In many cases, we \emph{can} remove the barrier and relax the outer loop dependency while still computing the correct result. How? Consider two tasks $A$ and $B$ that, in a BSP implementation, are ordered: $A$ is in an iteration that precedes $B$ and thus must run before $B$.
\begin{itemize}
  \item One possibility is to \emph{speculate} that we can compute $A$ and $B$ at the same time, or even in the order $B$ then $A$, without changing the correctness of the computation. If our speculation is correct, then we expose more concurrency. If our speculation is incorrect, then we must fix it. This fix might be cheap or costly.
  \item A second possibility is a problem formulation that is robust to computing items out of order. This is also called Dijkstra's don't care non-determinism~\cite{Dijkstra:1976:ADO}.
\end{itemize}
In either case, we can relax global barriers and expose additional concurrency; depending on the problem and dataset, we may see significant performance gains. However, relaxing barriers may incur additional costs: the cost of performing incorrectly speculated work, the cost of repairing incorrectly speculated work, and less predictable convergence rates when compared to the BSP counterpart. Overall, if the performance improvements from increased concurrency outweigh these costs, we can deliver performance gains.

\textbf{Related work:}
Hassaan et al.~\cite{Hassaan:2011:OVU} studied unordered and ordered versions of several algorithms, demonstrating a tradeoff between parallelism and work efficiency. However, the relaxed barrier formulations we study differ from unordered ones. Consider breadth-first search (BFS)\@. Both Hassaan et al.\ and we begin with a work-efficient Dijkstra BFS, but they compare to a work-inefficient Bellman-Ford BFS, while we compare to a relaxed (speculative) Dijkstra BFS (Section~\ref{sec:bfs-algorithm}). The speculative Dijkstra BFS is more work-efficient than Bellman-Ford BFS. Empirically, speculative Dijkstra's workload is within a small constant factor of that of BSP Dijkstra, which is $\# edges$ (see Table~\ref{tbl:extrawork}). This is much smaller than Bellman-Ford's workload of $diameter \times \#edges$.
Kulkarni et al.~\cite{kulkarni2007optimistic} studied an abstraction and runtime scheme for workloads with optimistic parallelism, which differ from the relaxing barriers we study in this paper.
Their notion of optimistic parallelism assumes many tasks can run in parallel and stops a task the moment it violates a dependency.
In contrast, we allow the computation to commit, even if it violates a dependency, and only fix the mistake afterwards.


\subsection{Worker Size}

Previous task-parallelism work~\cite{Cederman:2008:ODL, chen2010dynamic, augonnet2010data, rossbach2011ptask, krieder2014design, ben2017groute} uses the whole GPU as a single worker. They maintain a task dependency graph on the CPU side and orchestrate the execution by launching a CUDA kernel for each task when dependencies are satisfied. We use \emph{GPU-wide-worker} to describe such an organization. Tasks in those GPU-wide-worker task-parallel frameworks are usually very large to better utilize the entire GPU's resources. This organization is easy to program and has low scheduling overhead.

However, the GPU-wide-worker scheme is a poor match when task dependencies require finer management. Consider the following extreme example: Task $A$ and Task $B$ both contain 10,000 data items, and only a single data item in $B$ depends on a single item in $A$\@. A GPU-wide-worker implementation must wait for $A$ to complete before beginning work on $B$, even though most of the data items can be processed independently and concurrently. We use the term \emph{false dependency} to describe the situation when a data item has all its dependencies satisfied, but cannot be processed because another data item in the same task has unresolved dependencies. One can reduce such false dependencies by decomposing a large task into many smaller tasks to expose more parallelism. As a result, workers should be smaller, matching the size of tasks and allowing full utilization of the GPU's resources. This approach has motivated a number of recent task-parallelism frameworks~\cite{tzeng2010task, tzeng2012gpu, yeh2017pagoda, juggler, Steinberger:2014:WTS}, which use workers sized as either warps or Cooperative Thread Arrays (CTAs). The resulting additional complexity in scheduling many smaller workers motivates also moving scheduling decisions from the CPU to the GPU\@.

Most task-parallelism frameworks only provide one worker size. Our framework provides thread-, warp-, and CTA-sized workers, to support tasks of different size and different synchronization requirements. The only prior work that uses multiple granularities is Whippletree~\cite{Steinberger:2014:WTS}. Whippletree's thread and warp worker sizes are primarily a programming model concept and suffer from synchronization penalties at the implementation level. In Whippletree's implementation, threads are still synchronized within entire CTAs, suffering from false dependencies if tasks require finer synchronization than at CTA granularity.

\subsection{Balance Between Data and Task Parallelism}
\label{ss:mixture}

Many parallel applications have work items that require different amounts of processing. Traditional BSP applications address this challenge with explicitly coded data-parallel load balancing techniques. We describe two different approaches in the context of Listing 1:
One widely used technique is load balancing search~\cite{davidsonsssp}, which dynamically computes the prefix-sum of \verb|workload(i).size()| for $i \in$ \verb|in_frontier|, then flattens the two for-loops into one big array and redistributes the work in the array to each CUDA thread (see Baxter~\cite{lbs} for details). 
Another popular data-parallel load balancing technique separates the work in \verb|in_frontier| into different buckets based on \verb|workload(i).size()| and launch a separate kernel with the best processing strategy for that size for each bucket~\cite{merrillbfs}.

Task parallelism is a natural fit for these irregular applications. Workers in our framework do not directly synchronize with each other. One of the primary advantages of this lack of coordination is that it allows workers to attend to any available work items as soon as they become available
(``implicit task-parallel load balancing'').

Task parallelism and data parallelism are not exclusive-- individual tasks of sufficient size may also exploit data parallelism in their execution. Thus we consider a continuous spectrum with pure task-parallel and pure data-parallel load balancing at the extremes, and expect that the optimal trade-off will be application-dependent.
Our framework supports two worker sizes larger than a thread (warp and CTA) and offers the programmer the ability to exploit data parallelism within each warp-sized or CTA-sized task. In the framework, workers operate on tasks asynchronously, but an individual worker itself is executed synchronously. Therefore, we can use a worker's capacity as a parameter to control the tradeoff between data and task parallelism. Given a fixed number of threads available, increasing a worker's capacity reduces the total number of workers available for a given application. At the same time, an increase in worker capacity results in more opportunities to perform data-parallel load balancing within each worker. In the extreme, setting a worker's capacity to the entire GPU leaves no room for task parallelism and is equivalent to the BSP model. We found that data-parallel load balancing inside the worker combined with task-parallel load balancing provided by Atos results in better overall load balancing (Section~\ref{sec:analysis}). We are not aware of any previous work that combines these two types of load balancing.
\vspace{-10pt}
\subsection{Kernel Strategy}
Traditional GPU kernels divide a variable amount of input work into fixed-size CTAs and launch a kernel over a CTA count proportional to the amount of input work. Persistent kernels~\cite{Gupta:2012:ASO} decouple the relationship between data size and launched CTAs. A persistent kernel launches only enough CTAs to fill the GPU\@. These CTAs remain resident for the entire kernel and run a loop that matches naturally to the task-parallelism model in Listing~\ref{lst:tp}.

\textbf{Advantages of persistent kernels} Persistent kernels reduce kernel launch overhead and CPU/GPU communication. This is particularly significant where many small kernels are required. The persistent kernel reduces CPU involvement in favor of programmer-written GPU logic within the persistent kernel.

\textbf{Disadvantages of persistent kernels} GPU workers in the persistent kernel concurrently pop from a shared queue; this requires atomic operations to ensure exclusive pops. Persistent kernels have higher register usage than discrete kernels (requiring extra registers to maintain the queue loop).

Intuitively, if a discrete-kernel application suffers from large kernel overhead (exhibits the small frontier problem), a persistent kernel may be preferred. Otherwise, it may be better and simpler to choose a discrete kernel. Previous task-parallelism frameworks either use discrete kernels~\cite{Cederman:2008:ODL,augonnet2010data, rossbach2011ptask, krieder2014design})  or persistent kernels~\cite{tzeng2010task, tzeng2012gpu, yeh2017pagoda, juggler, Steinberger:2014:WTS}, but none of them provide both and/or expose that decision to the programmer. In Section~\ref{sec:kernel-strategy-results}, we discuss the results of our experiments with respect to this choice.

%% file: tex/framework_api.tex
\section{Framework API} \label{s:api}
With the discussion from Section~\ref{sec:design_decisions} in mind, we introduce the Atos API\@ shown in Listing~\ref{listing:framework_api}.
\vspace{-10pt}
\begin{lstfloat}
\begin{code}
\begin{minted}{cuda}
template<typename T, typename COUNTER_T>
struct Queues {
    __host__ void init (COUNTER_T, int num_queues, int iteration);

    template<typename F1, typename F2, typename... Args>
    __host__ void launchThread (bool ifPersist, int numBlock, 
    int numThread, int shareMem_size, F1 f1, F2 f2, Args... arg);

    template<typename F1, typename F2, typename... Args>
    __host__ void launchWarp (bool ifPersist, int numBlock, 
    int numThread, int sharedMem_size, F1 f1, F2 f2, Args... arg);

    template<int FETCH_SIZE, typename F1, typename F2, typename... Args>
    __host__ void launchCTA (bool ifPersist, int numBlock, 
    int numThread, int shareMem_size, F1 f1, F2 f2, Args... arg);
}

\end{minted}
\caption{Atos framework APIs}
\label{listing:framework_api}
\end{code}
\end{lstfloat}
\verb|launch*| API functions are used to launch workers who repeatedly pop tasks from the work queue; each worker then applies function \verb|f1| to the task popped.  When the worker fails to pop, it runs function \verb|f2| (default noop) instead. Under persistent kernel mode, \verb|numThread|$\times$\verb|numBlock| cannot exceed the maximum number of threads that can concurrently reside on the GPU  based on the application's register and shared memory usage. By default, these are set to the maximum allowed values, but can be overridden by users.
As discussed in Section~\ref{sec:design_decisions}, when we execute \verb|launchCTA|, the choice of \verb|numThread| will determine the fraction of task- and data-parallelism. Argument \verb|FETCH_SIZE| defines how many data items constitute a task to be popped from the queue by a worker. Given \verb|numBlock|, the \verb|FETCH_SIZE| should be set accordingly. Increasing \verb|FETCH_SIZE| will reduce available tasks for other workers but increase local data parallelism, and decreasing \verb|FETCH_SIZE| will have the opposite effect.

%% file: tex/graph-problems.tex
\section{Three Case Studies}
\label{sec:graph-problems}
In this paper we select three classic nested loop problems in the domain of graph computation for in-depth study. We choose them because their implementations are well-studied in the BSP model so we can be confident that our results are meaningful.
Also when run on particular datasets, their BSP implementations exhibit one or more of the challenges described in Section~\ref{sec:bsp-performance-challenges}. In Algorithms 1--6, we use CUDA atomic operation APIs~\cite{cuda_api}.

\subsection{Breadth-First Search}
\label{sec:bfs-algorithm}

The pseudocodes of BSP BFS and relaxed-barrier BFS (``Speculative BFS'') are shown in Algorithms~\ref{alg:bsp_bfs} and~\ref{alg:relax_bfs}. Algorithm~\ref{alg:bsp_bfs} shows that BSP BFS is exactly Dijkstra's algorithm: BSP BFS advances the outer iteration only when all vertices in the  \verb|in_frontier| are traversed; thus strict breadth-first ordering is preserved, ensuring every vertex will first be reached via its optimal path.

In contrast, in Algorithm~\ref{alg:relax_bfs}, Speculative BFS relaxes the barrier constraints of the outer loop of BSP BFS\@.  Speculative BFS has a single dynamic frontier, and many independent CUDA workers asynchronously push and pop vertices to the frontier. Consequently, vertices at different distances can be processed simultaneously. This raises the possibility of extra work, since out-of-order iteration may require visiting vertices multiple times in order to find the shortest path.

Despite this difference, we stress that both algorithms are based on a similar vertex traversal.
Speculative BFS is more precisely seen as Dijkstra with relaxed barrier constraints, and differs from SSSP algorithms such as Bellman-Ford. 

\subsection{PageRank}
PageRank computes the importance (rank) of nodes in a graph.
PageRank has a ``push'' and a ``pull'' formulation. Both BSP and relaxed-barrier PageRank (asynchronous PageRank) use push and their pseudocodes are shown in Algorithm~\ref{alg:bsp_pr} and~\ref{alg:relax_pr}. The algorithm begins with an initial rank and residue value per node. In Algorithm~\ref{alg:bsp_pr}, BSP PageRank is computed by two kernels that are repeatedly called in an outer loop until all vertices converge. The first kernel pushes a fraction of the residue of each vertex on the \verb|frontier| to its neighbors. The second kernel aggregates all vertices with residue $> \epsilon$ and adds them to the \verb|frontier|. Having two separate kernels allows threads to be remapped for more flexible load balancing.

Asynchronous PageRank (Algorithm~\ref{alg:relax_pr}) removes the global synchronization and fuses the two kernels from the BSP implementation together. In asynchronous PageRank, CUDA workers pop vertices from the queue asynchronously, push their residue to their neighbors, then exclusively reserve \verb|Check_Size| number of vertices and push those vertices onto the queue if their residue $> \epsilon$.


PageRank is an iterative unordered algorithm as it is essentially a random walk~\cite{chung2010pagerank,fortunato2007random}. In theory, the BSP global barrier is not necessary or beneficial. In our experiments, BSP PageRank does not converge faster than asynchronous PageRank.

\subsection{Graph Coloring}
\label{ss:gc_alg}
Graph coloring is usually implemented using either an indep\-en\-dent-set-based approach or a greedy approach. In this paper, \emph{both} the BSP and relaxed barrier graph coloring use a speculative greedy graph coloring algorithm~\cite{gebremedhin2000scalable} shown in Algorithms~\ref{alg:bsp_gc} and~\ref{alg:relax_gc}.
In Algorithm~\ref{alg:bsp_gc}, the BSP implementation uses a double buffer and consists of two kernels that are called in the outer loop until all vertices are appropriately colored. The first kernel assigns colors to vertices from \verb|in_frontier| while avoiding color conflicts with their neighbors. The second kernel aggregates each vertex that, after the first kernel's color assignment, still has a color conflict with a neighbor, and adds it to \verb|out_frontier|. The speculative part of this algorithm is in the first kernel: it allows a vertex to be colored using the outdated color information from its neighbors and if the vertex color assignment fails (and is detected in the second kernel), that vertex is re-added to the frontier for recoloring.

In contrast, in Algorithm~\ref{alg:relax_gc}, the relaxed barrier version fuses the two kernels by an uberkernel~\cite{hargreaves2005generating}; we use the sign of the vertex ID to distinguish between the color assignment task and the conflict detection task. Specifically, if we pop a vertex with positive ID, we perform color assignment (lines 8--11 of Algorithm~\ref{alg:relax_gc}, identical to lines 11--14 of Algorithm~\ref{alg:bsp_gc}), and if we pop a vertex with negative ID, we perform conflict detection (lines 16--18 of Algorithm~\ref{alg:relax_gc}, identical to lines 21--23 of Algorithm~\ref{alg:bsp_gc}). Thus, Algorithm~\ref{alg:relax_gc}  is almost identical to Algorithm~\ref{alg:bsp_gc}. The only difference is that Algorithm~\ref{alg:relax_gc} does not enforce a global barrier between the color assignment and conflict detection kernels.
\vspace{-3pt}
\begin{algorithm}

    \begin{minted}
      [
      % frame=lines,
      framesep=1mm,
      baselinestretch=1.2,
      fontsize=\scriptsize,
      xleftmargin=1em,linenos,
      baselinestretch=1,
      %highlightlines={8-12},
      % fontsize=8,
      % linenos
      ]
      {python}
# Initialization
Graph G
in_frontier = [source]
for vertex in G:
  vertex.dist = MAX_UINT32
# Start BFS
while not in_frontier.empty():
  out_frontier = []
  -------------CUDA_KERNEL--------------
  for vertex in in_frontier:
     for neighbor in vertex.neighbors:
       neighber_dist=atomicMin(&neighbor.dist, vertex.dist+1)
       if vertex.dist + 1 < neighbor_dist:
          out_frontier.append(neighbor)
  -----------------------------------------
  cudaDeviceSynchronize()
  in_frontier = out_frontier
  \end{minted}
  \caption{Bulk Synchronous BFS}
  \label{alg:bsp_bfs}
\end{algorithm}
\vspace{-12pt}

\begin{algorithm}
    \begin{minted}
      [
      % frame=lines,
      framesep=1mm,
      baselinestretch=1.2,
      fontsize=\scriptsize,
      xleftmargin=1em,linenos,
      baselinestretch=1,
      % fontsize=8,
      % linenos
      ]
      {python}
#Same initialization as Algorithm 1
#Start BFS
-------------CUDA_KERNEL----------------
while not frontier.empty():
  vertex = frontier.pop()
  for neighbor in vertex.neighbors:
    neighber_dist=atomicMin(&neighbor.dist, vertex.dist+1)
    if vertex.dist + 1 < neighbor_dist:
      frontier.append(neighbor)
----------------------------------------
    \end{minted}
\caption{Speculative BFS}
\label{alg:relax_bfs}
\end{algorithm}

\begin{algorithm}
    \begin{minted}
      [
      % frame=lines,
      framesep=2mm,
      baselinestretch=1.2,
      fontsize=\scriptsize,
      xleftmargin=1em,linenos,
      baselinestretch=1,
      % fontsize=8,
      % linenos
      ]
      {python}
# Initialization
Graph G
rank[G.total_vertices] = {1-lambda}
residue[G.total_vertices] = {0}
for vertex in G.all_vertices():
  atomicAdd(residue+vertex, (1-lambda)/lambda*vertex.neighborLen)
  frontier.append(vertex)
# Start PageRank
while not frontier.empty():
  -------------CUDA_KERNEL----------------
  for vertex in frontier:
    res = atomicExch(residue+vertex,0)
    rank[vertex] = rank[vertex]+res
    res = res*lambda/vertex.neighborLen
    for neighbor in vertex.neighbors:
      atomicAdd(residue+neighbor, res)
  ----------------------------------------
  cudaDeviceSynchronize()
  frontier = []
  -------------CUDA_KERNEL----------------
  for vertex in G.all_vertices():
    if residue[vertex] > epsilon:
      frontier.append(vertex)
  ----------------------------------------
  cudaDeviceSynchronize()
    \end{minted}
\caption{Bulk Synchronous PageRank}
\label{alg:bsp_pr}
\end{algorithm}
\vspace{-10pt}

\begin{algorithm}
    \begin{minted}
      [
      % frame=lines,
      framesep=2mm,
      baselinestretch=1.2,
      fontsize=\scriptsize,
      xleftmargin=1em,linenos,
      baselinestretch=1,
      % fontsize=8,
      % linenos
      ]
      {python}
#Same initialization as Algorithm 3
#Start PageRank
-------------CUDA_KERNEL-----------------------
while not frontier.empty():
  vertex = frontier.pop()
  res = atomicExch(residue+vertex,0)
  atomicAdd(rank+vertex, res)
  res = res*lambda/vertex.neighborLen
  for neighbor in vertex.neighbors:
    atomicAdd(residue+neighbor, res)

  check_start = atomicAdd(check, Check_Size)
  for check_id in check_start+Range[0,Check_Size):
     if residue[check_id%G.total_vertices] > epsilon:
        frontier.append(check_id%G.total_vertices)
  ----------------------------------------------
    \end{minted}
\caption{Asynchronous PageRank}
\label{alg:relax_pr}
\end{algorithm}
\vspace{-10pt}

\begin{algorithm}
    \begin{minted}
      [
      % frame=lines,
      framesep=2mm,
      baselinestretch=1.2,
      fontsize=\scriptsize,
       xleftmargin=1em,linenos,
       baselinestretch=1,
      % fontsize=8,
      % linenos
      ]
      {python}
# Initialization
Graph G
in_frontier = G.all_vertices()
for vertex in in_frontier:
  vertex.color = -1;
# Start graph coloring
while not in_frontier.empty():
  # Assign each vertex a color different from neighbors'
  -------------CUDA_KERNEL-----------------------
  for vertex in in_frontier:
    forbidden[G.max_degree] = {0}
    for neighbor in vertex.neighbors:
      forbidden[neighbor.color] = 1
    vertex.color = find_min_color(forbidden[color] == false)
  ----------------------------------------------
  cudaDeviceSynchronize()
  # Collect vertices whose colors collide with neighbors'
  out_frontier = []
  -------------CUDA_KERNEL-----------------------
  for vertex in in_frontier:
    for neighbor in vertex.neighbors:
      if vertex.color == neighbor.color:
        out_frontier.append(vertex)
  ----------------------------------------------
  cudaDeviceSynchronize()
  in_frontier = out_frontier
    \end{minted}
\caption{Bulk Synchronous speculative Graph Coloring}
\label{alg:bsp_gc}
\end{algorithm}
\vspace{-1pt}

\begin{algorithm}
    \begin{minted}
      [
      % frame=lines,
      framesep=2mm,
      baselinestretch=1.2,
      fontsize=\scriptsize,
       xleftmargin=1em,linenos,
      baselinestretch=1,
      % fontsize=8,
      % linenos
      ]
      {python}
#Same initialization as Algorithm 5
#Start graph coloring
-------------CUDA_KERNEL-----------------------
while not frontier.empty():
  # Assign each vertex a color different from neighbors'
  vertex = frontier.pop()
    if vertex > 0
      forbidden[G.max_degree] = {0}
      for neighbor in vertex.neighbors:
        forbidden[neighbor.color] = 1
      vertex.color = find_min_color(forbidden[color] == false)
      frontier.append(-1*vertex)
   # push vertices whose colors collide with neighbors'
    else if vertex < 0
      vertex = -1 * vertex
      for neighbor in vertex.neighbors:
        if vertex.color == neighbor.color:
          frontier.append(vertex)
----------------------------------------------
    \end{minted}
\caption{Asynchronous speculative Graph Coloring}
\label{alg:relax_gc}
\end{algorithm}

%% file: tex/analysis.tex
\section{Experiments and Analysis} \label{sec:analysis}
\subsection{System and Hardware Information}
All experiments in this paper are run on a Linux workstation with 2.20~GHz Intel(R) hyper-threaded E5-2698 v4 Xeon(R) CPUs, 528~GB of main memory, and an NVIDIA V100 GPU with 32~GB on-board memory. All programs were compiled with NVIDIA's nvcc compiler (version 11.1.168) with the -O3 flag and gcc 9.3.0 with the -O3 flag. All results ignore transfer time and are averaged over 20 runs.

\subsection{Experimental Overview}
We evaluate the design principles discussed in Section~\ref{sec:design_decisions} using three implementation variants based on a combination of Atos configurations. (1)~``persist-32'' utilizes persistent kernels with \emph{warp-sized} workers. It has no data-parallel load balancing within a worker, instead only using implicit task-parallel load balancing. (2)~``persist-worker\_size-FETCH\_SIZE'' utilizes persistent kernels with \emph{CTA-sized} workers. (3)~``discrete-worker\_size-FETCH\_SIZE'' utilizes discrete kernels and \emph{CTA-sized} workers. Both CTA variants use load balancing search~\cite{merrillbfs} (a data-parallel load balancing technique) inside workers in conjunction with implicit task-parallel load balancing.

For BFS and PageRank, we compare the performance of our implementations to Gunrock~\cite{gunrock2017}, a state-of-the-art single-GPU BSP-based graph framework, which extensively uses data-parallel load-balancing techniques. For Graph Coloring, Gunrock's independent-set graph coloring algorithm is not comparable, so we faithfully implemented a BSP graph coloring using the same speculative greedy graph coloring algorithm. Our BSP implementation uses the Gunrock's bucket-based data-parallel load balancing method~\cite{gunrock2017}, described in Section \ref{ss:mixture}.


We run the three case studies on three scale-free and two mesh-like datasets (see Table~\ref{tbl:dataset}) and summarize the runtime (speedup) results for four implementations in Table~\ref{tbl:perf}.

\begin{table}
  \centering
  \caption{Runtime with speedup comparing to BSP in the paraphrase of BSP and three Atos implementations. Graph types are s (scale-free) and m (mesh-like). Runtime unit: ms}
  \label{tbl:perf}
  \scriptsize
  \begin{tabular}{*{5}{p{.1\linewidth}}}
    \multicolumn{5}{c}{Application: BFS  (BSP=Gunrock)} \\
    \toprule
    \multicolumn{1}{p{.05\linewidth}}{Dataset} &
                                                 \multicolumn{1}{p{.05\linewidth}}{BSP} &
                                                                                          \multicolumn{1}{p{.1\linewidth}}{\begin{tabular}{@{}c@{}}persist\\warp\end{tabular}} &
                                                                                                                                                                                 \multicolumn{1}{p{.1\linewidth}}{\begin{tabular}{@{}c@{}}persist\\CTA\end{tabular}}
                                               & \multicolumn{1}{p{.1\linewidth}}{\begin{tabular}{@{}c@{}}discrete\\CTA\end{tabular}} \\

    \midrule
    \multicolumn{1}{l}{soc-LiveJournal1\textsuperscript{s}} & \multicolumn{1}{l}{15.3} & \multicolumn{1}{l}{22.3  (x0.68)} & \multicolumn{1}{l}{12.4  (x1.23)} & \multicolumn{1}{l}{10.7  (x1.42)} \\
    \multicolumn{1}{l}{hollywood\_2009\textsuperscript{s}} & \multicolumn{1}{l}{9.26} & \multicolumn{1}{l}{12.2 (x0.75)} & \multicolumn{1}{l}{6.23 (x1.48)} & \multicolumn{1}{l}{4.56 (x2.02)} \\
    \multicolumn{1}{l}{indochina\_2004\textsuperscript{s}} & \multicolumn{1}{l}{13.2} & \multicolumn{1}{l}{15.6 (x0.84)} & \multicolumn{1}{l}{8.03 (x1.65)} & \multicolumn{1}{l}{7.42 (x1.79)}\\
    \multicolumn{1}{l}{road\_usa\textsuperscript{m}} & \multicolumn{1}{l}{604} & \multicolumn{1}{l}{327 (x1.84)} & \multicolumn{1}{l}{46.9 (x12.8)} & \multicolumn{1}{l}{174 (x3.46)} \\
    \multicolumn{1}{l}{roadNet\_ca\textsuperscript{m}} & \multicolumn{1}{l}{55.9} & \multicolumn{1}{l}{39.6 (x1.41)} & \multicolumn{1}{l}{4.35 (x12.8)} & \multicolumn{1}{l}{15.5 (x3.58)} \\
    \bottomrule
    \multicolumn{5}{c}{Application: PageRank (BSP=Gunrock)} \\
    \toprule
    \multicolumn{1}{p{.05\linewidth}}{Dataset} &
                                                 \multicolumn{1}{p{.05\linewidth}}{BSP} &
                                                                                          \multicolumn{1}{p{.1\linewidth}}{\begin{tabular}{@{}c@{}}persist\\warp\end{tabular}} &
                                                                                                                                                                                 \multicolumn{1}{p{.1\linewidth}}{\begin{tabular}{@{}c@{}}persist\\CTA\end{tabular}}
                                               & \multicolumn{1}{p{.1\linewidth}}{\begin{tabular}{@{}c@{}}discrete\\CTA\end{tabular}} \\

    \midrule
    \multicolumn{1}{l}{soc-LiveJournal1\textsuperscript{s}} & \multicolumn{1}{l}{262} & \multicolumn{1}{l}{156 (x1.68)} & \multicolumn{1}{l}{113 (x2.31)} & \multicolumn{1}{l}{116 (x2.25)} \\
    \multicolumn{1}{l}{hollywood\_2009\textsuperscript{s}} & \multicolumn{1}{l}{87.1} & \multicolumn{1}{l}{80.0 (x1.08)} & \multicolumn{1}{l}{68.5 (x1.27)} & \multicolumn{1}{l}{72.4 (x1.20)} \\
    \multicolumn{1}{l}{indochina\_2004\textsuperscript{s}} & \multicolumn{1}{l}{159} & \multicolumn{1}{l}{84.7 (x1.88)} & \multicolumn{1}{l}{52.6 (x3.02)} & \multicolumn{1}{l}{49.6 (x3.20)}\\
    \multicolumn{1}{l}{road\_usa\textsuperscript{m}} & \multicolumn{1}{l}{221} & \multicolumn{1}{l}{169 (x1.30)} & \multicolumn{1}{l}{121 (x1.81)} & \multicolumn{1}{l}{112 (x1.95)} \\
    \multicolumn{1}{l}{roadNet\_ca\textsuperscript{m}} & \multicolumn{1}{l}{20.5} & \multicolumn{1}{l}{16.2 (x1.26)} & \multicolumn{1}{l}{10.1 (x2.03)} & \multicolumn{1}{l}{8.28  (x2.47)} \\
    \bottomrule
    \multicolumn{5}{c}{Application: Graph Coloring} \\
    \toprule
    \multicolumn{1}{p{.05\linewidth}}{Dataset} &
                                                 \multicolumn{1}{p{.05\linewidth}}{BSP}  &
                                                                                           \multicolumn{1}{p{.1\linewidth}}{\begin{tabular}{@{}c@{}}persist\\warp\end{tabular}} &
                                                                                                                                                                                  \multicolumn{1}{p{.1\linewidth}}{\begin{tabular}{@{}c@{}}persist\\CTA\end{tabular}}
                                               & \multicolumn{1}{p{.1\linewidth}}{\begin{tabular}{@{}c@{}}discrete\\warp\end{tabular}}  \\
    \midrule
    \multicolumn{1}{l}{soc-LiveJournal1\textsuperscript{s}} & \multicolumn{1}{l}{96.5} & \multicolumn{1}{l}{20.4  (x4.71)} & \multicolumn{1}{l}{36.1 (x2.67)} & \multicolumn{1}{l}{63.2  (x1.52)} \\
    \multicolumn{1}{l}{hollywood\_2009\textsuperscript{s}} & \multicolumn{1}{l}{77.9} & \multicolumn{1}{l}{31.9 (x2.4)} & \multicolumn{1}{l}{59.3 (x1.31)} & \multicolumn{1}{l}{274 (x0.28)} \\
    \multicolumn{1}{l}{indochina\_2004\textsuperscript{s}} & \multicolumn{1}{l}{673} & \multicolumn{1}{l}{74.1 (x9.08)} & \multicolumn{1}{l}{184 (x3.65)} & \multicolumn{1}{l}{2073 (x0.32)}\\
    \multicolumn{1}{l}{road\_usa\textsuperscript{m}} & \multicolumn{1}{l}{38.2} & \multicolumn{1}{l}{51.4 (x0.74)} & \multicolumn{1}{l}{19.3 (x1.97)} & \multicolumn{1}{l}{81.9 (x0.46)} \\
    \multicolumn{1}{l}{roadNet\_ca\textsuperscript{m}} & \multicolumn{1}{l}{9.11} & \multicolumn{1}{l}{4.18 (x2.18)} & \multicolumn{1}{l}{3.52 (x2.58)} & \multicolumn{1}{l}{12.0  (x0.75)} \\
    \bottomrule
  \end{tabular}
\end{table}
\begin{table}
  \scriptsize
   \caption{Summary of datasets used in our experiments. Graph types are s (scale-free) and m (mesh-like)}
   \label{tbl:dataset}
    \begin{tabular}{lrrrrrr}
    \toprule
    &  &  &  & Max. &  Max.  & Avg. \\
    Dataset & Vertices & Edges & Diam. & indeg. & outdeg.  & degree \\
    \midrule
    soc-LiveJournal1\textsuperscript{s} & 4.8M & 68M & 20 & 13,905 & 20,292 & 14 \\
    hollywood\_2009\textsuperscript{s} & 1.1M & 11M & 11 & 11,467 & 11,467 & 105 \\
    indochina\_2004\textsuperscript{s} & 7.4M & 191M & 26 & 256,425 & 6,984 & 8 \\
    road\_usa\textsuperscript{m} & 23.9M & 57M & 6,809 & 9 & 9 & 2 \\
    roadNet\_ca\textsuperscript{m} & 1.9M & 5M & 849 & 12 & 12 & 2 \\
    \bottomrule
  \end{tabular}
\end{table}

\subsection{Performance Challenges in Three Study Cases}
\label{ss:characteristics}
Each of our study cases embodies a subset of the BSP performance challenges discussed in Section~\ref{sec:bsp-performance-challenges}, which informs our choice of design decisions in the Atos framework. We summarize the BSP performance challenges in Table~\ref{tbl:characteristics_of_case}.

\textbf{Load Imbalance Problem:} All three algorithms involve iterating over a vertex's neighbor list; thus variance in the vertices' degrees leads to load imbalance. This issue is much more severe on scale-free datasets where vertex degree variance is high. In contrast, mesh-like graphs have low maximum degree, and hence low degree variance (see Table~\ref{tbl:dataset}).

\textbf{Small Frontier Problem:} In Figures~\ref{fig:bfs_thp_time}, \ref{fig:pr_thp_time}, and~\ref{fig:gc_thp_time}, we plot throughput against timeline for each BSP implementation of the three algorithms. Low throughput over a long duration of time indicates the presence of the small-frontier problem.

\emph{BFS:} In Figure~\ref{fig:bfs_thp_time}, Gunrock has high throughput on scale-free datasets, and thus does not have the small-frontier problem. This is because scale-free datasets have low diameter (leading to a small number of BSP iterations) and high average degree (leading to a large amount of work per iteration). In contrast, Gunrock on mesh-like datasets \emph{does exhibit} the small frontier problem, because these datasets have high diameters and small average degree; consequently, there is a large number of iterations, with little work per iteration, leading to low throughput over many iterations.

\emph{PageRank:} In Figure~\ref{fig:pr_thp_time}, for Gunrock PageRank, both scale-free and mesh-like datasets do not exhibit the small frontier problem, as they have high throughput over most of the execution time and converge in fewer than 35 iterations (though Indochina-2004 exhibits a long flat tail in the latter half of execution).

\emph{Graph Coloring:} In Figure~\ref{fig:gc_thp_time}, for BSP graph coloring, scale free datasets have low throughput for more than 70\% of execution time, and thus have the small frontier problem. Mesh-like datasets terminate in fewer than 40 iterations, and have short tails, and thus do not have the small frontier problem. This is because on scale-free datasets, the high-degree vertices will need to be recolored many times, leading to a large number of iterations, during which the frontier contains a few high-degree vertices with color conflicts. In contrast, mesh-like datasets have low average degree, and are less likely to have color conflicts.
\begin{table}
  \centering
  \caption{Summary of performance challenges for each case study.}
  \label{tbl:characteristics_of_case}
  \scriptsize
  \begin{tabular}{lccc}
    \toprule
    & BFS &  PageRank  & Graph Coloring \\
    \midrule
    Scale-Free & Load Imbalance & Load Imbalance & Load Imbalance + Small Frontier\\
    Mesh-Like & Small Frontier & None & None\\
    \bottomrule
  \end{tabular}
\end{table}

\subsection{Relaxing Barriers}

\begin{table}
  \centering
  \caption{Upper: Workload ratio of three Atos implementations relative to Gunrock's implementations for BFS and PageRank. A workload ratio of $n$ means that our implementation does $n$ times as much work as Gunrock. Lower: the workload ratio of four implementations relative to the input graph's total vertex count for graph coloring.}
  \label{tbl:extrawork}
  \scriptsize

  \begin{tabular}{*{7}{p{.1\linewidth}}}
    \toprule
    \multicolumn{1}{p{.08\linewidth}}{} & \multicolumn{3}{c|}{Application: BFS} & \multicolumn{3}{c}{Application: PageRank} \\
    \midrule
    \multicolumn{1}{p{.08\linewidth}}{Dataset}
                                        & \multicolumn{1}{p{.08\linewidth}}{ \begin{tabular}{@{}c@{}}persist\\warp\end{tabular} }
                                        & \multicolumn{1}{p{.08\linewidth}}{\begin{tabular}{@{}c@{}}persist\\CTA\end{tabular}}
                                        & \multicolumn{1}{c|}{\begin{tabular}{@{}c@{}}discrete\\CTA\end{tabular}}
                                        & \multicolumn{1}{p{.08\linewidth}}{ \begin{tabular}{@{}c@{}}persist\\warp\end{tabular} }
                                        & \multicolumn{1}{p{.08\linewidth}}{\begin{tabular}{@{}c@{}}persist\\CTA\end{tabular}}
                                        & \multicolumn{1}{p{.08\linewidth}}{\begin{tabular}{@{}c@{}}discrete\\CTA\end{tabular}} \\
    \midrule
    \multicolumn{1}{l}{soc-LiveJournal1\textsuperscript{s}} & \multicolumn{1}{l}{1.43} & \multicolumn{1}{l}{1.06} &  \multicolumn{1}{l}{1.01}
                                        & \multicolumn{1}{c}{0.73} & \multicolumn{1}{c}{0.72} & \multicolumn{1}{c}{0.72} \\
    \multicolumn{1}{l}{hollywood\_2009\textsuperscript{s}} & \multicolumn{1}{l}{2.26} & \multicolumn{1}{l}{1.19} &  \multicolumn{1}{l}{1.07}
                                        & \multicolumn{1}{c}{1.08} & \multicolumn{1}{c}{1.18} & \multicolumn{1}{c}{0.9} \\
    \multicolumn{1}{l}{indochina\_2004\textsuperscript{s}} & \multicolumn{1}{l}{1.28} & \multicolumn{1}{l}{1.00} &  \multicolumn{1}{l}{1.00}
                                        & \multicolumn{1}{c}{0.76} & \multicolumn{1}{c}{0.73} & \multicolumn{1}{c}{0.75} \\
    \multicolumn{1}{l}{road\_usa\textsuperscript{m}} &\multicolumn{1}{l}{3.56} & \multicolumn{1}{l}{1.05} &  \multicolumn{1}{l}{1.04}
                                        & \multicolumn{1}{c}{0.79} & \multicolumn{1}{c}{0.79} & \multicolumn{1}{c}{0.92} \\
    \multicolumn{1}{l}{roadNet\_ca\textsuperscript{m}} &  \multicolumn{1}{l}{2.05} & \multicolumn{1}{l}{1.02} & \multicolumn{1}{l}{1.04}
                                        & \multicolumn{1}{c}{1.18} & \multicolumn{1}{c}{1.11} & \multicolumn{1}{c}{0.97} \\
    \bottomrule
  \end{tabular}

  \begin{tabular}{*{5}{p{.1\linewidth}}}
    \multicolumn{5}{c}{Application: Graph Coloring} \\
    \toprule
    \multicolumn{1}{p{.2\linewidth}}{Dataset} &  \multicolumn{1}{p{.13\linewidth}}{BSP} &
                                                                                          \multicolumn{1}{p{.13\linewidth}}{ \begin{tabular}{@{}c@{}}persist\\warp\end{tabular} } &
                                                                                                                                                                                    \multicolumn{1}{p{.13\linewidth}}{\begin{tabular}{@{}c@{}}persist\\CTA\end{tabular}}
                                              & \multicolumn{1}{p{.13\linewidth}}{\begin{tabular}{@{}c@{}}discrete\\warp\end{tabular}} \\
    \midrule
    \multicolumn{1}{l}{soc-LiveJournal1\textsuperscript{s}} & \multicolumn{1}{l}{1.17} & \multicolumn{1}{l}{1.00} & \multicolumn{1}{l}{1.74} & \multicolumn{1}{l}{2.78} \\
    \multicolumn{1}{l}{hollywood\_2009\textsuperscript{s}} & \multicolumn{1}{l}{3.31} & \multicolumn{1}{l}{1.15} & \multicolumn{1}{l}{5.24} & \multicolumn{1}{l}{\textcolor{blue}{37.34}} \\
    \multicolumn{1}{l}{indochina\_2004\textsuperscript{s}} & \multicolumn{1}{l}{1.96} & \multicolumn{1}{l}{1.04} & \multicolumn{1}{l}{4.45} & \multicolumn{1}{l}{\textcolor{blue}{16.97}} \\
    \multicolumn{1}{l}{road\_usa\textsuperscript{m}} & \multicolumn{1}{l}{1.22} & \multicolumn{1}{l}{1.00} & \multicolumn{1}{l}{1.46} & \multicolumn{1}{l}{1.41} \\
    \multicolumn{1}{l}{roadNet\_ca\textsuperscript{m}} & \multicolumn{1}{l}{2.55} & \multicolumn{1}{l}{1.00} & \multicolumn{1}{l}{1.74} & \multicolumn{1}{l}{2.44} \\
    \bottomrule
  \end{tabular}
\end{table}
As discussed in Sections~\ref{sec:problem-class}--\ref{sec:design_decisions}, relaxing barriers exposes more concurrency, giving higher throughput and shorter execution time. However, relaxing barriers may result in extra work. If the performance improvement from increased concurrency outweighs the cost of extra work, we obtain a net performance gain.

There are two key factors influencing this tradeoff. First, we find that in the presence of a small frontier problem, the increase in concurrency from relaxing barriers is always more significant than the cost of extra work. Second, on naturally unordered algorithms such as PageRank, one can always relax the barrier: although the barrier gives the BSP implementation a more predictable convergence rate, the barrier generally does not make the convergence faster.

\john{I don't like this second conclusion. It's not helpful. Obviously if the barrier is unimportant (PageRank) for correctness AND convergence, we can remove it. I'd like more interesting conclusions here. For, say, BFS on mesh vs.\ scale-free networks, or graph coloring on those, which has more overwork and why? How much does the cost of repair influence the decision of relaxing the barrier? Those are interesting conclusions.}

\begin{figure}
  \centering
  \includegraphics[width=\linewidth]{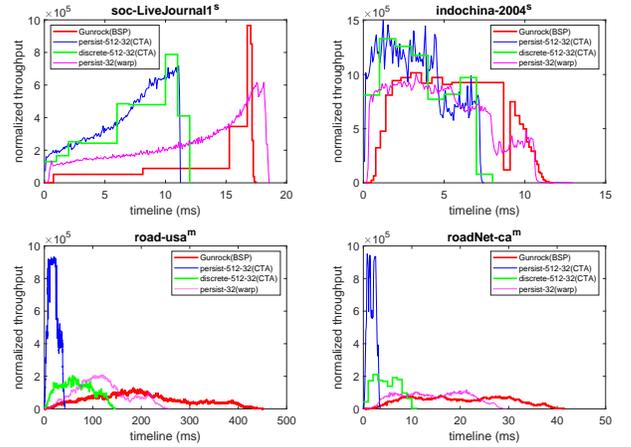}
  \caption{Normalized throughput vs.\ time on BFS\@. The top charts are scale-free; bottom charts are mesh-like.}
  \label{fig:bfs_thp_time}
\end{figure}

\begin{figure}
  \centering
  \includegraphics[width=\linewidth]{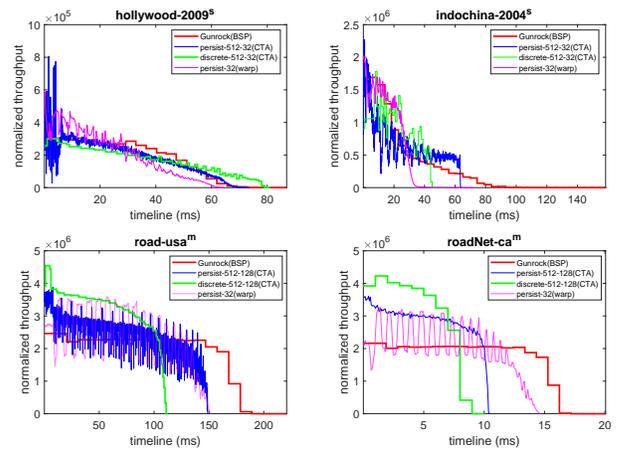}
  \caption{Normalized throughput vs.\ time on PageRank.}
    \vspace{-10pt}
  \label{fig:pr_thp_time}
\end{figure}

\begin{figure}
  \centering
  \includegraphics[width=\linewidth]{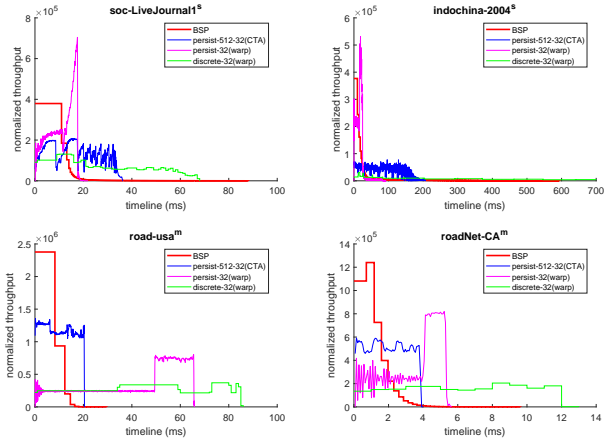}
  \caption{Normalized throughput vs.\ time on graph coloring.}
  \vspace{-10pt}
  \label{fig:gc_thp_time}
\end{figure}

Table~\ref{tbl:extrawork} summarizes the extra work for three study cases. Figures~\ref{fig:bfs_thp_time},~\ref{fig:pr_thp_time} and~\ref{fig:gc_thp_time} plot the throughput of four implementations (BSP~+ three Atos variants) of three study cases against timeline for four datasets. Notably, these plots show the \emph{normalized throughput}, which is the measured throughput divided by the overwork factors in Table~\ref{tbl:extrawork}. This gives a fair measure of overall performance, as it incorporates both the benefits of improved concurrency (higher absolute throughput) and the cost of extra work. Essentially, normalized throughput measures ``useful'' throughput rather than raw absolute throughput. We provide detailed analysis below for each application.

\textbf{BFS:} Figure~\ref{fig:bfs_thp_time} shows that for the two mesh-like datasets, all 3 Atos implementations achieve considerably higher normalized throughput than Gunrock. Why? Table~\ref{tbl:characteristics_of_case} shows that Gunrock on mesh-like datasets has a severe small frontier problem. Therefore, the increase in concurrency in the 3 Atos implementations offers a significant performance advantage. Table~\ref{tbl:extrawork} indicates that the persistent-warp implementation generates 3.5x extra work vs.\ Gunrock, but despite this extra work, Atos's normalized throughput is still significantly higher than Gunrock.  Scale-free graphs, on the other hand, exhibit more parallelism and do not suffer from the small-frontier problem. Atos's fastest implementations are still faster than Gunrock's, but not nearly as much as for the mesh networks. 

On all BFS experiments, Atos's CTA implementations are faster than its warp ones. 
Atos's CTA implementations use a combination of task-parallel and data-parallel load balancing techniques (see Section~\ref{ss:worker_size} for details), and thus have better load balancing than its warp implementations, which only use task-parallel load balancing. This leads to higher GPU utilization and hence higher absolute throughput. Second, CTA implementations produce less extra work than warp (see Table~\ref{tbl:extrawork}). 
Due to better load balancing in CTA, the workload of each worker has lower variance. If a worker receives too much work, there will be a long delay before the vertices' updated depths are visible to other workers; this increases the likelihood that downstream vertices are first reached via other sub-optimal paths, which leads to extra work.

\textbf{PageRank:} Unlike BFS, PageRank is naturally unordered, as it satisfies Dijkstra's don't care non-determinism~\cite{Dijkstra:1976:ADO}. Therefore, relaxing the barrier in the outer loop does not generate any misspeculations and hence results in no wasted work. In fact, Table~\ref{tbl:extrawork} shows that the Atos implementations perform \emph{less} work than Gunrock in general. This is because the BSP barrier forces each vertex to be processed at most once per iteration. By relaxing this barrier, the Atos implementations can update certain important vertices (e.g., vertices with high centrality) more frequently than other vertices, thus leading to more efficient propagation of rank.

Figure~\ref{fig:pr_thp_time} shows that all three Atos implementations compact the workload and process it with higher normalized throughput (persist-CTA has a higher profiling cost).
\john{The previous sentence has a leap in logic---``higher throughput is because of no global barrier''---that isn't really justified by this figure.}
Though PageRank does not suffer from the small frontier problem, the three Atos implementations nonetheless have superior performance over Gunrock, because relaxing the barrier increases concurrency. In addition, relaxing the barrier lowered the overall workload in practice, even though in theory it may lead to a more unpredictable convergence rate.



\textbf{Graph Coloring:}
Unlike BFS and PageRank, all graph coloring implementations (including BSP) use a speculative approach (greedy graph coloring) and thus all have extra work. Table~\ref{tbl:extrawork} summarizes the multiplicative factor of extra work, which is defined as a ratio vs.\ the number of vertices in the graph (the lowest possible workload). Atos's persist warp has the least extra work; on some datasets, the extra work is less than 1\%, which means after the first color assignment, only 1\% of vertices have a color conflict and must be recolored.
Atos's discrete warp has the most extra work (on hollywood-2009, 37.34x). The extra work is due to the combination of two factors:

\textit{1. Conflicts tend to arise when neighboring vertices are colored concurrently:} From Section~\ref{ss:gc_alg}, given a vertex, the algorithm first checks its neighbors' colors, then assigns a color to the vertex that does not conflict with its neighbors. The color assignment is speculative because it is done using possibly outdated color information from the vertex's neighbors. When neighboring vertices are colored simultaneously, they read outdated colors from each other, leading to conflicts and recoloring.

\emph{2. Consecutive vertices on the work queue are likely to be neighbors:} On many if not most graphs, the vertex ID is semantically meaningful: vertices whose vertex ID are numerically close are more likely to be neighbors. At the beginning of graph coloring, all vertices are initially inserted onto the work queue in order of vertex ID\@.

\label{sec:why-random-permutation-of-vertices}
Since consecutive vertices on the work queue tend to be assigned colors concurrently, the above implies a high likelihood of color conflicts. We verify that the large amount of extra work is indeed due to semantically meaningful vertex IDs: running the exact same experiment with randomly permuted vertex IDs, the amount of extra work drops to less than 1.5x for all four implementations on all datasets. ID permutation leads to the following runtime improvements (in ms) on scale-free datasets:

\vspace{-1mm}
\noindent
\begin{center}
\begin{small}
\begin{tabular}{lccc}
  \toprule
  Impl. & soc-LiveJournal1 & hollywood & indochina\\
  \midrule
  discrete-warp & 63 $\rightarrow$ 31 & 274 $\rightarrow$ 26 & 2073 $\rightarrow$ 222 \\
  persist-CTA & 36 $\rightarrow$ 21 & 59 $\rightarrow$ 28 & 184 $\rightarrow$ 50 \\
  BSP & 96 $\rightarrow$ 89 & 77 $\rightarrow$ 61 & 673 $\rightarrow$ 485 \\
  \bottomrule
\end{tabular}
\end{small}
\end{center}
\vspace{-1mm}

\noindent
The BSP implementation has a more modest improvement because BSP's thread-warp-CTA load balancing scheme~\cite{merrillbfs} already divides each bucket into three individually-load-balanced subbuckets, reducing inter-bucket conflicts.
Persist-warp has little change as there is almost no extra work even before permutation. Notably, after permutation, all three Atos variants are faster than BSP implementation on scale-free datasets.

\textbf{Comparing persist-warp and persist-CTA:} persist-CTA has better load balancing, allowing for more (potentially adjacent) vertices to be colored simultaneously, resulting in more extra work than persist-warp. We verify this from Table~\ref{tbl:extrawork}.
Roughly speaking, the amount of extra work for persist-CTA is more significant on scale-free graphs, as a vertex can have a large number of neighbors, leading to more potential conflicts. Therefore, persist-CTA outperforms persist-warp on mesh-like graphs, where the increased concurrency and better load-balancing outweigh the cost of wasted work; conversely, on scale-free datasets, persist-CTA is slower than persist-warp, because the cost of extra work is too high (see Tables~\ref{tbl:perf} and~\ref{tbl:extrawork}).

\john{I don't see why the last point is relevant. You're not comparing with a BSP implementation; both persist-warp and persist-CTA don't care about small-frontier.}

\textbf{Comparing persist-warp and discrete-warp:} Discrete-warp has more extra work than persist-warp, hurting its performance, for two reasons.

1)~The scheduling policies of discrete and persistent kernels are different. When kernels are launched from the CPU (discrete-kernel strategy), the kernel launched earlier always has a higher scheduling (hardware) priority than the kernel launched later. This effectively causes vertices to always be colored in roughly the same order as their initial ordering (by vertex ID, which causes many conflicts). In contrast, the persistent kernel only incurs one kernel launch and warps within it are scheduled by the hardware scheduler, whose decisions are much less ordered by vertex ID\@. Thus persist-warp has fewer coloring conflicts caused by adjacencies and hence less overwork.

2)~discrete-warp has lower register usage than persistent-warp (72 vs.\ 42), so persist-warp only achieves 43\% occupancy per SM and discrete-warp achieves 62\%. Therefore the discrete-wrap assigns colors to more vertices simultaneously, leading to a greater likelihood of conflicts than persist-warp. Unlike our other applications, in graph coloring, the cost of extra work largely reduces the benefit of increased concurrency. On scale-free datasets (without random permutation), our highest performance is achieved with lower concurrency and less overwork (persist-warp variant), which achieves a lower absolute throughput but a higher normalized throughput (and hence higher performance overall).

\subsection{Worker Size and Trade-off between Task- and Data-Parallelism Load Balancing}
\label{ss:worker_size}
\begin{figure}
  \centering
  \includegraphics[width=\linewidth, angle=0]{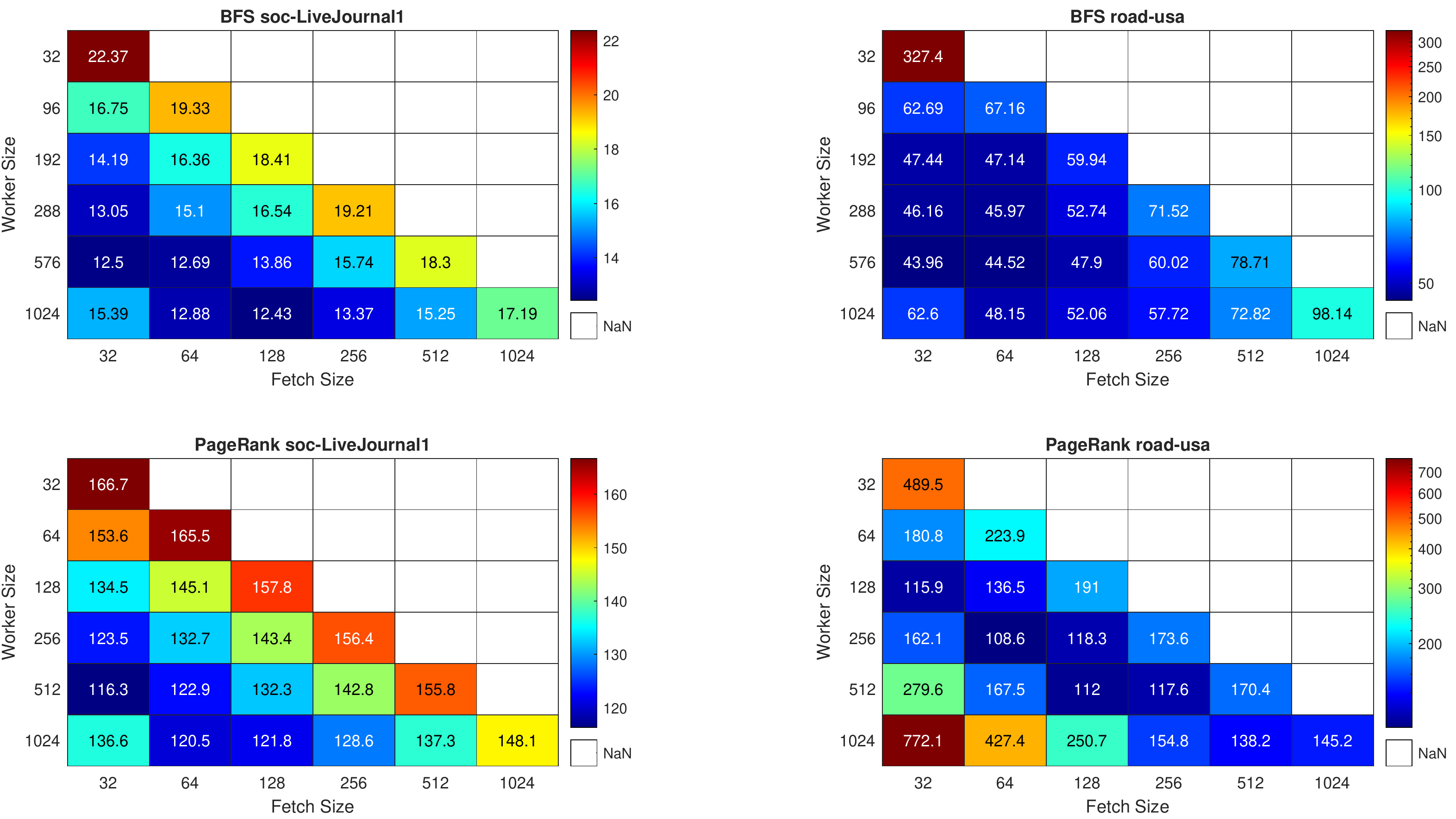}
  \caption{Runtime (ms) heatmap plotted with different worker size and fetch size for BFS and PageRank on soc-LiveJournal\textsuperscript{s} and road\_usa\textsuperscript{m}. Note only the lower triangle is valid.}
  \label{fig:heatmap}
\end{figure}

As discussed in Section~\ref{sec:design_decisions}, Atos enables the user to trade off between task-parallelism and data-parallelism load balancing by adjusting the worker size and FETCH\_SIZE\@. Atos persist-CTA\@, like persist-warp, uses a persistent kernel to exploit task parallelism, but now the task-parallel work units are fewer and larger (the size of a CTA) and we can leverage more data parallelism within a CTA\@. In most cases, persist-CTA outperforms persist-warp with both higher normalized/absolute throughput, except for the graph coloring on scale-free datasets, where it achieves only higher absolute throughput. Figure~\ref{fig:heatmap} illustrates this tradeoff for BFS and PageRank on soc-LiveJournal (scale-free)  and road\_usa (mesh-like). We exclude graph coloring because it can only be run with one CTA size, due to high register usage (72) and high shared memory usage (46~KB).


\vspace{-0.5\baselineskip}

\subsection{Kernel Strategy}
\label{sec:kernel-strategy-results}
From Section~\ref{sec:design_decisions}, the chief advantage of the persistent kernel is removing the overhead associated with kernel invocation, which is most significant for fine-grained tasks that involve many small kernel launches. Based on the performance results in Table~\ref{tbl:perf} and Figure~\ref{fig:bfs_thp_time}, the performance gap between persistent kernel and discrete kernel is particularly large for BFS on mesh-like graphs, which require many small kernel launches due to the high diameter and small workload per iteration. Graph coloring on indochina-2004 also shows a large kernel launch overhead. Using a random permutation of vertex IDs (see Section~\ref{sec:why-random-permutation-of-vertices}), Atos's persistent variant is 4.3x faster than its discrete variant.

%% file: tex/conclusion.tex
\vspace{-0.5\baselineskip}
\section{Conclusion}
\label{sec:conclusion}
In this paper, we present our task-parallel GPU dynamic scheduling framework, Atos, and analyze its performance across numerous design parameters on three case studies. Our analysis provides the following guidelines on what applications are suitable to run in a capable task-parallel framework, as well as what Atos configurations to use, given an application's characteristics:
\begin{enumerate}
\item If the dynamic application either exhibits the small frontier problem or has load imbalance, Atos will have a performance advantage.
\item If the application exhibits the small frontier problem, it should be run with a persistent kernel.
\item If the application exhibits load imbalance, it should be run with both task- and data-parallelism load balancing in tandem to achieve better performance. For different applications, the optimal tradeoff point varies.
\item By relaxing the outer loop dependency in the application, Atos increases concurrency at the cost of extra work due to mis-speculation, or less predictable convergence rates. The optimal tradeoff between the increased concurrency and additional cost is application-dependent. When an application is naturally unordered (e.g., PageRank) or  has the small frontier problem (e.g., BFS on mesh-like datasets and graph coloring on scale-free datasets), the increased concurrency usually outweighs the cost. Conversely, on problems such as BFS on scale-free graphs or graph coloring on mesh-like graphs, the cost of extra work can hurt performance. The best way to reduce extra work is application-dependent and may include better load balancing (e.g., BFS) or reducing concurrency (e.g., graph coloring).
\end{enumerate}